\documentclass[9pt,twocolumn,twoside]{optica}
\newcommand{\cmmnt}[1]{}
\setboolean{shortarticle}{true}
\setboolean{minireview}{false}
\usepackage{xcolor}
\usepackage{notes2bib}
\title{Phase stabilization of a coherent fibre network by single-photon counting}

\author[1,2,*]{Salih Yanikgonul}
\author[1]{Ruixiang Guo}
\author[3]{Angelos Xomalis}
\author[1]{Anton N. Vetlugin}
\author[1]{Giorgio Adamo}
\author[1]{Cesare Soci}
\author[1,3]{Nikolay I. Zheludev}

\affil[1]{Centre for Disruptive Photonic Technologies, School of Physical and Mathematical Sciences and The Photonics Institute, Nanyang Technological University, 637371, Singapore}
\affil[2]{Institute of Materials Research and Engineering, Agency for Science, Technology and Research, 138632, Singapore}
\affil[3]{Optoelectronics Research Centre and Centre for Photonic Metamaterials, University of Southampton, Southampton SO17 1BJ, United Kingdom}

\affil[*]{Corresponding author: salih001@ntu.edu.sg}

\begin{abstract}
Coherent optical fibre networks are extremely sensitive to thermal, mechanical and acoustic noise, which requires elaborate schemes of phase stabilization  with dedicated auxiliary lasers, multiplexers and photodetectors. This is particularly demanding in quantum networks operating at the single-photon level. Here we propose a simple method of phase stabilization based on single-photon counting and apply it to quantum fibre networks implementing single-photon interference on a lossless beamsplitter and coherent perfect absorption on a metamaterial absorber. As a proof of principle, we show dissipative single-photon switching with visibility close to 80\%. This method can be employed in quantum networks of greater complexity without classical stabilization rigs, potentially increasing efficiency of the quantum channels.
\end{abstract}

\setboolean{displaycopyright}{true}

\begin{document}

\maketitle

Leveraging on advanced telecommunication technologies, coherent optical fibre networks provide a scalable platform for quantum light processing and quantum communication. Single-photon interference~\cite{Xavier:11,Cho:09}, quantum computation based on a dual-rail qubit encoding~\cite{Lukens:17,Lu:18}, quantum key distribution~\cite{Li:18,Liu:18}, entanglement swapping and distribution~\cite{Sun:17,Ikuta2018} have already been demonstrated in fibre environment proving feasibility of fiber-based quantum optics. Recent achievements in fabrication of fully-fiberized metamaterial packages~\cite{Roger2015} allow to extend the functionality of accessible devices for quantum light manipulation. For instance, quantum coherent perfect absorption (CPA) with plasmonic metamaterial absorber, first demonstrated in a free space~\cite{PhysRevLett.117.023601}, was shown in an optical fibre network~\cite{Anton:19}. However, to deal with phase noise – an inherent problem of fibre systems, these experiments had to employ resource demanding stabilization or elaborate data post selection techniques~\cite{Xavier:11,Cho:09,Cho:16,Anton:19}.

In this paper, we develop a new method of phase stabilization by single-photon counting which is far less resource demanding, and apply it to a basic coherent network represented by a fully-fiberized Mach-Zehnder interferometer (MZI) operating at the single-photon level. We further implement this technique in a coherent optical fibre network containing plasmonic metamaterial absorber operating at CPA regime, where we achieve deterministic control of single-photon absorption probability. We show all-optical switching by driving the system between absorbing and transmitting modes, otherwise not possible in the previous experiment without stabilization~\bibnote[Anton:19]{A. N. Vetlugin, R. Guo, A. Xomalis, S. Yanikgonul, G. Adamo, C. Soci and N. I. Zheludev, "Coherent perfect absorption of single photons in a fibre network," Appl. Phys. Lett, doc. ID APL19-AR-06056R1 (posted 23 October 2019, in press).}

\begin{figure}[t]
\includegraphics[width=\columnwidth]{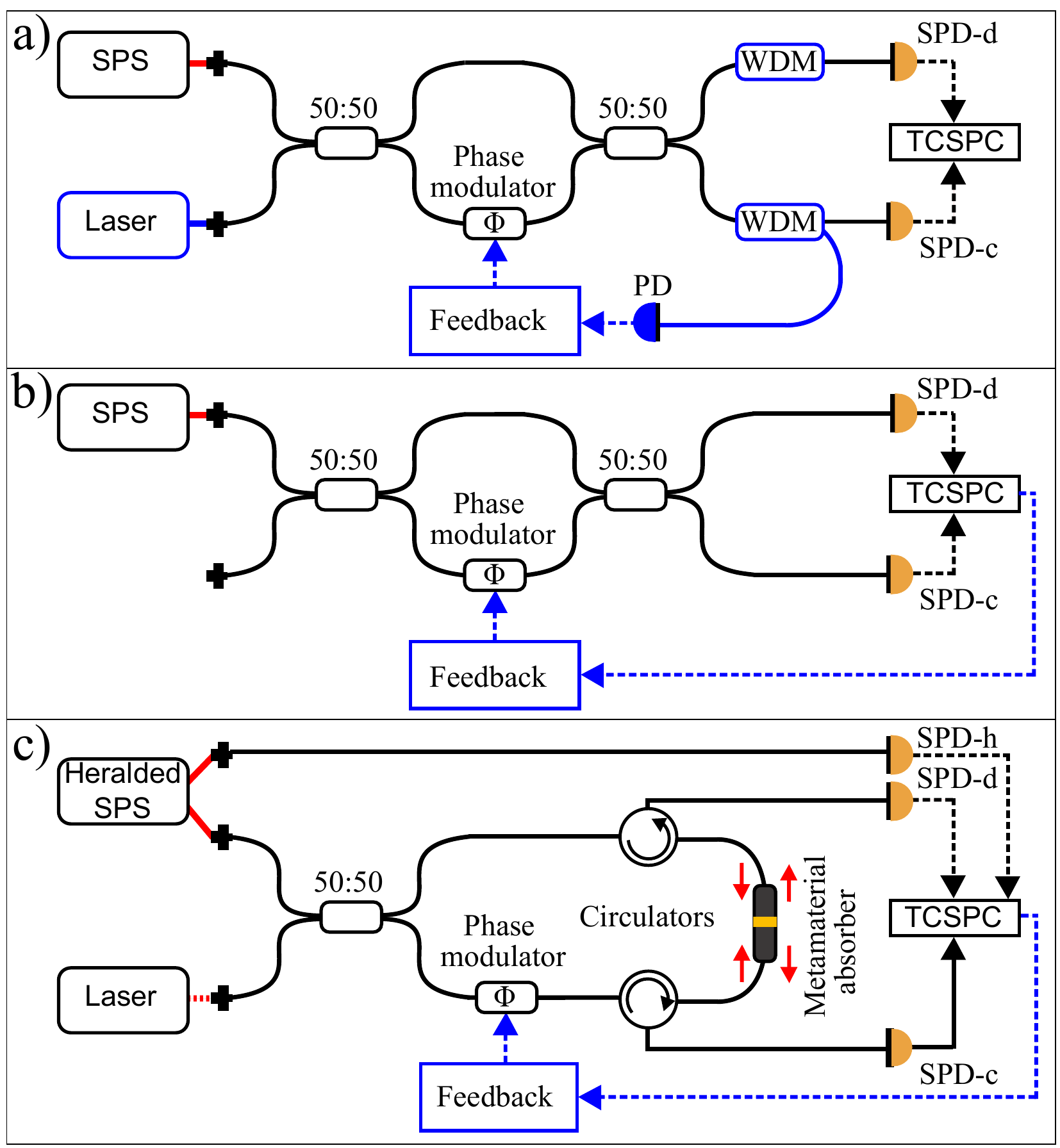}
\caption{\small Stabilization of coherent optical fibre networks operating at the single-photon level. a) A MZI consisting of a single-photon source (SPS), 50:50 BSs and a phase modulator as an example network. Time-correlated single-photon counting (TCSPC) module processes signals of single-photon detectors (SPD). In conventional stabilization schemes (highlighted in blue), a laser light sent through the coherent network is separated and measured by a multiplexer (e.g. WDM) and a photodetector (PD), respectively. b) Optical fibre networks stabilized with single-photons: TCSPC output is used as an input of the feedback system. c) Single-photon absorption control and switching experiment setup, where the second BS is substituted with a metamaterial absorber.
}
\label{fig:1}
\end{figure}

\begin{figure}[t]
\centering
\includegraphics[width=0.965\columnwidth]{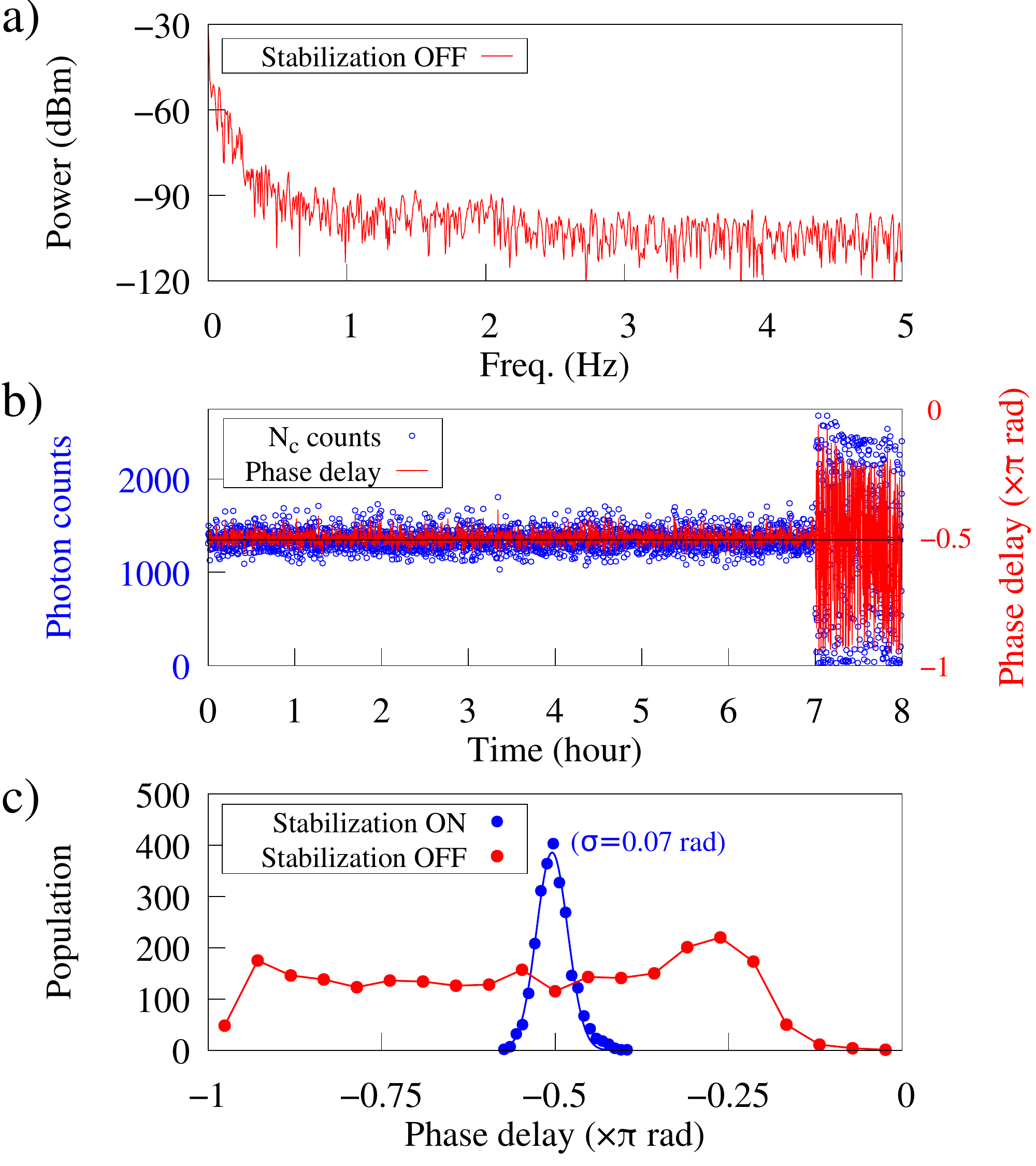}\\
\caption{\label{fig:2}\small Long-term phase stabilization in a fully-fiberized MZI. a) The noise spectral density of phase fluctuations of the unstabilized system, measured by sending a CW-laser of $\mu$W power through the same interferometer. b) The long-term phase stabilization where the system is stabilized for 7 hours, followed by another 7 hours without stabilization (only one hour is shown). Each blue circle corresponds to single-photon counts with the red line being the corresponding phase delay retrieved from these counts. The black line represents the stabilization point. c) The corresponding phase distributions for the stabilized (blue) and unstabilized (red) periods, where the blue line is the Gaussian fit curve.
}
\end{figure}

The extreme sensitivity of optical fibre networks to thermal, mechanical and acoustic noise~\cite{Musha:82} strongly affects coherence and indistinguishability of quantum light. One way to overcome this problem is employing a feedback loop for an active phase stabilization, which requires an auxiliary laser with a distinct degree of freedom (e.g. wavelength, polarization or temporal mode) probing the same optical path traversed by quantum light~\cite{Xavier:11,Cho:09,Cho:16}. Fig.~\ref{fig:1}\textcolor{blue}{(a)} shows an example of a fiberized implementation of the single-photon MZI consisting of 50:50 beamsplitters (BS) and a phase modulator backbone, and the auxiliary phase stabilization loop. Phase fluctuations affect the intensity distribution of the light detected at the output ports of the MZI. To compensate this effect, the intensity of the auxiliary laser propagating through the MZI is used to generate a feedback signal sent to the phase modulator. The dedicated multiplexers (e.g. wavelength division multiplexers, dichroic mirrors, and polarizers), filters and photodetectors are required to isolate the feedback circuit and the quantum channel. This unavoidably degrades the quality of the quantum signal.

Alternatively, single-photons themselves can be used to accomplish the same task, Fig.~\ref{fig:1}\textcolor{blue}{(b)}. In the single-photon regime, the photon wavefunction acquires different phases while passing through different arms of the MZI, which affects the photon distribution at the two output ports. The very same output of the single-photon detectors (SPDs) can be used to generate the feedback signal based on the expected photon rate, and the experiment can subsequently be performed within a time window shorter than the characteristic time of the phase fluctuation. The cost for this simplicity compared to stabilization schemes based on auxiliary components is 1) lower frequency of operation as single-photon counting requires comparatively long integration time, 2) discontinuous mode of operation, since experimental measurements are interleaved with stabilization periods. 

In this letter, first, we introduce the new phase stabilization scheme. Next, the experimental results obtained from a fully-fiberized conventional MZI stabilized with this technique is presented to show the applicability of the method to the fiberized quantum optics experiments. We conclude the paper with a demonstration of a practical application, a dissipative single-photon switch, in a quantum fibre network, which is operated at the regime of CPA and stabilized by employing this method.

To assess the phase stabilization by single-photon counting, we performed a basic experiment with fully-fiberized MZI at the single-photon level~(Fig.~\ref{fig:1}\textcolor{blue}{(b)}). In this first experiment, a strongly attenuated CW-laser is used as a single-photon source; photon wavefunction is split on a first 50:50 BS into a superposition of two spatial modes, corresponding to the upper and lower arms of the interferometer (each arm is composed of around 15 meters of polarization-maintaining single-mode fibres). Due to the interference on the second BS, probabilities to detect a photon by SPD-c ($p_\textrm{c}$) and SPD-d ($p_\textrm{d}$) depend on the phase retardation $\phi$ between two optical paths,~\cite{Anton:19} 
\begin{equation}
  \label{eq:BS-p}
  \begin{aligned}
    p_\textrm{c}(\phi) = (1 + \sin(\phi))/2\\
    p_\textrm{d}(\phi) = (1 - \sin(\phi))/2
  \end{aligned}
\end{equation}

To estimate these probabilities, the number of counts of SPD-c ($N_\textrm{c}$) and SPD-d ($N_\textrm{d}$) should be measured during the time interval $\Delta t$ smaller than the characteristic time of the phase noise,

\begin{equation}
  \label{eq:BS-N}
  \begin{aligned}
    N_\textrm{c}(\phi) = N \cdot p_\textrm{c}(\phi)\\
    N_\textrm{d}(\phi) = N \cdot p_\textrm{d}(\phi)\\
  \end{aligned}
\end{equation}

\setlength{\parindent}{0ex}where $N$ is the number of photons arriving at the second BS during $\Delta t$. In our setup, the noise is present on a scale of $\Delta f_\textrm{N}\approx$1~Hz without significant contributions above this frequency (Fig.~\ref{fig:2}\textcolor{blue}{(a)}), for which we set $\Delta t=24$~ms ($\Delta t \ll 1/\Delta f_\textrm{N}$).

\setlength{\parindent}{3ex}Since $\phi$ is the only parameter that determines the detected photon numbers $N_\textrm{c}$ and $N_\textrm{d}$, we may monitor the phase stability of the network by measuring the variation of SPDs counts. After scanning the phase to set $N_\textrm{c}$ at approximately $N/2$ level, we start the stabilization procedure to keep $N_\textrm{c}$ at the same level. If no significant phase fluctuations are present during $\Delta t$, then $N_\textrm{c}$ does not change (more precisely, number of detected photons would be within the Poisson distribution centered at $N/2$, considering the random nature of the laser source), and no action is required. Opposite, if the interferometer is affected by the noise during $\Delta t$, then the difference between detected counts $N_\textrm{c}$ and expected $N/2$ becomes noticeable. In this case, a feedback signal is applied to the phase modulator in order to compensate fluctuations, minimizing the absolute difference between detected and expected photon numbers.

The continuous stabilization of this method has been tested up to 7 hours (Fig.~\ref{fig:2}\textcolor{blue}{(b)}). Each blue circle in Fig.~\ref{fig:2}\textcolor{blue}{(b)} shows the number of single-photons detected by the SPD-c during $\Delta t$ with a sampling period of 10 seconds, while the red line represents the phase delay retrieved from these counts. The corresponding distribution for  the unstabilized and the stabilized periods are shown in Fig.~\ref{fig:2}\textcolor{blue}{(c)} together with the Gaussian fit-curve with the rms width of $\sigma = 0.07$~rad for the latter. This implies that the length difference between arms of the interferometer is kept within 10 nm range for 810 nm input photons, which correspond to a 9 orders of magnitude spatial resolution for 15 meters of fibres. This value is comparable to those demonstrated in conventional stabilization schemes~\cite{Xavier:11,Cho:09,Cho:16}, which demonstrates the possibility of controlling single-photons by single-photon counting in coherent optical fibre networks. We note that the stabilization of longer interferometers with enlarged noise bandwidth may require decrease of the integration time $\Delta t$ and increase of the single-photon source brightness.

\begin{figure}
\includegraphics[width=\columnwidth]{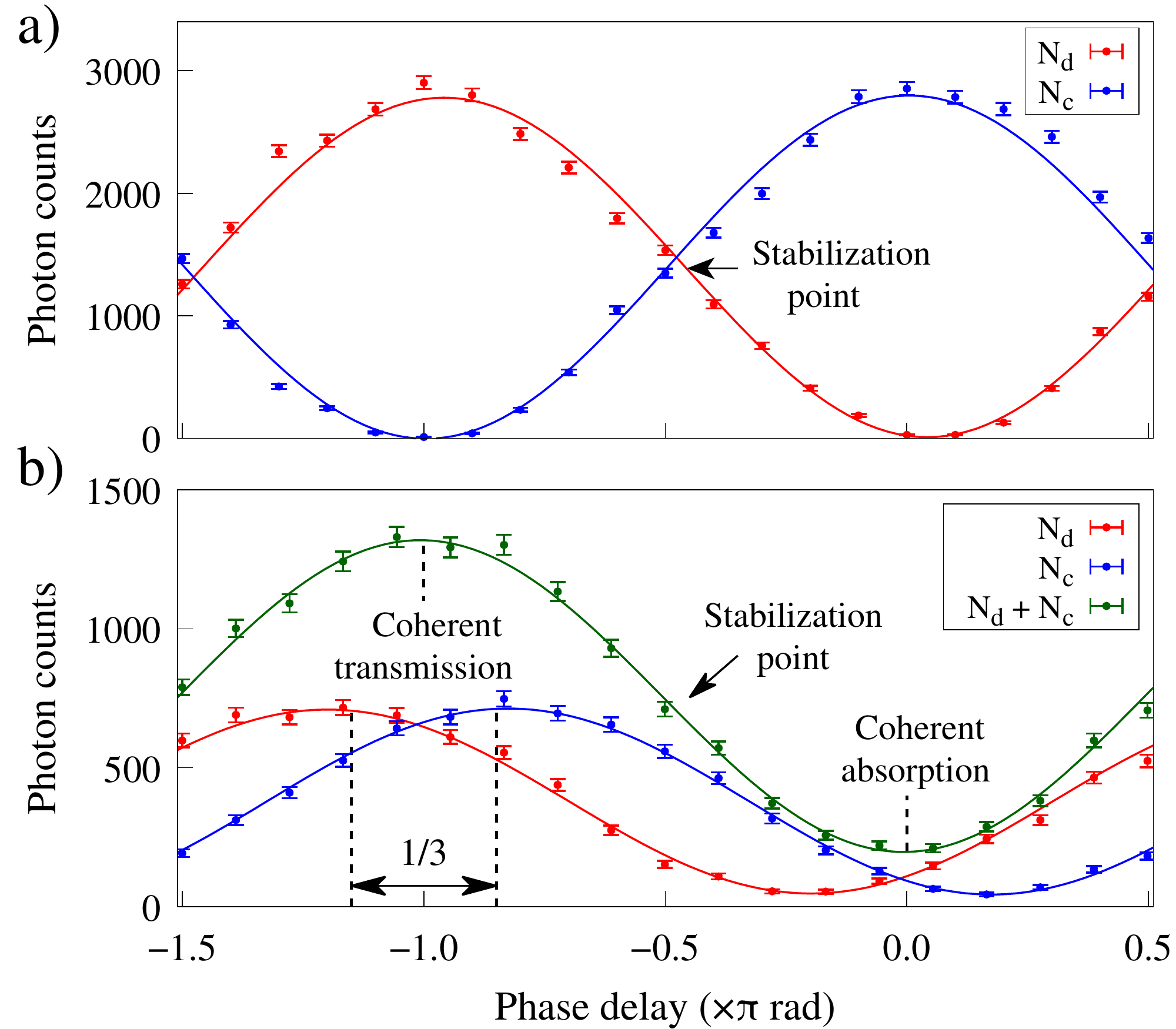}\\
\vspace*{-0.5cm}
\caption{\label{fig:3}\small Single-photon manipulation in coherent optical fibre networks stabilized by single-photons. a) Single-photon interference in a fully-fiberized MZI~(Fig.~\ref{fig:1}\textcolor{blue}{(b)}). Out-of-phase oscillation of $N_\textrm{c}$ and $N_\textrm{d}$ is in a good agreement with~Eqs.~(\ref{eq:BS-p},\ref{eq:BS-N}). b) Single-photon absorption control with a CPA~(Fig.~\ref{fig:1}\textcolor{blue}{(c)}). As each point corresponds to a single measurement, the dispersion is defined by the Poisson distribution.
}
\end{figure}

Next, to prove the applicability of this new technique to fiberized quantum optics experiments, we measured single-photon interference fringes according to~Eqs.~(\ref{eq:BS-p},\ref{eq:BS-N}), Fig.~\ref{fig:3}\textcolor{blue}{(a)}, in the stabilized MZI~(Fig.~\ref{fig:1}\textcolor{blue}{(b)}). The $N_\textrm{c}$ is first set at $N/2$ level and the system is stabilized at this point, which requires to set some phase retardation $\phi_\textrm{st}$ by the phase modulator. Then, the feedback is switched off, and the phase retardation is shifted to one from a discrete set,

\begin{equation}
\label{eq:phi-n}
\phi_\textrm{n} = \phi_\textrm{st} + n \cdot \Delta \phi
\end{equation}
\setlength{\parindent}{0ex}where $\Delta \phi \approx0.1 \pi$ and $n=0,\pm1,\pm2,...$ during time interval $\Delta t$. The corresponding photon number detected by SPD-c and SPD-d, $N_\textrm{c}(\phi_\textrm{n})$  and $N_\textrm{d}(\phi_\textrm{n})$, is recorded. After that, the phase retardation is stabilized back to $\phi_\textrm{st}$. In this way, a $2\pi$~phase spectrum is sampled. In addition to this static stabilization in which the set point is fixed (i.e. $N/2$), we note that the proposed technique can be extended to dynamical stabilization, performed at any point of the SPD-c curve without any need to back to $\phi_\textrm{st}$ after each measurement, or to Pound–Drever–Hall type of phase control~\cite{doi:10.1119/1.1286663} by periodically modulating phase retardation with an extra phase modulator.

\setlength{\parindent}{3ex}The experimental data~(Fig.~\ref{fig:3}\textcolor{blue}{(a)}) clearly demonstrates out-of-phase oscillation of $N_\textrm{c}$ and $N_\textrm{d}$ as it is expected according to theoretical predictions of~Eqs.~(\ref{eq:BS-p},\ref{eq:BS-N}). This verifies an ability to manipulate quantum light in coherent quantum network without a necessity to complicate the setup. 

Lastly, we show that a phase stabilized coherent fibre network operating at CPA regime can be used to control the single-photon absorption probability for coherent switching application. The CPA takes place in a setup shown in Fig.~\ref{fig:1}\textcolor{blue}{(c)}, when the output 50:50 BS is replaced by a coherent perfect absorber. This lossy component can be described by a four-port device with 

\begin{equation}
\label{eq:CPA-parameters}
 t = \pm r = 1/2
\end{equation}

\setlength{\parindent}{0ex}where $t$ and $r$ are amplitude transmission and reflection coefficients. In contrast to the ~Eq.~(\ref{eq:BS-p}), the photon detection probability at output ports of the absorber in this case is defined as~\cite{Anton:19}

\begin{equation}
\label{eq:CPA-p}
p_\textrm{c}(\phi) = p_\textrm{d}(\phi) = (1 - \cos \phi)/4
\end{equation}

\setlength{\parindent}{0ex}with a total absorption probability, $1-(p_\textrm{c}+p_\textrm{d})$, varying in the range from 0 to 1. Recently the CPA at the single-photon level was demonstrated in a fully-fiberized optical fibre network~\cite{Anton:19}, using advanced data post selection to overcome the phase noise. This data post selection technique, however, did not allowed to determine the absorption probability of single-photons on demand, which our new technique is capable of. 

\begin{figure}[t]
\centering
\includegraphics[width=\columnwidth]{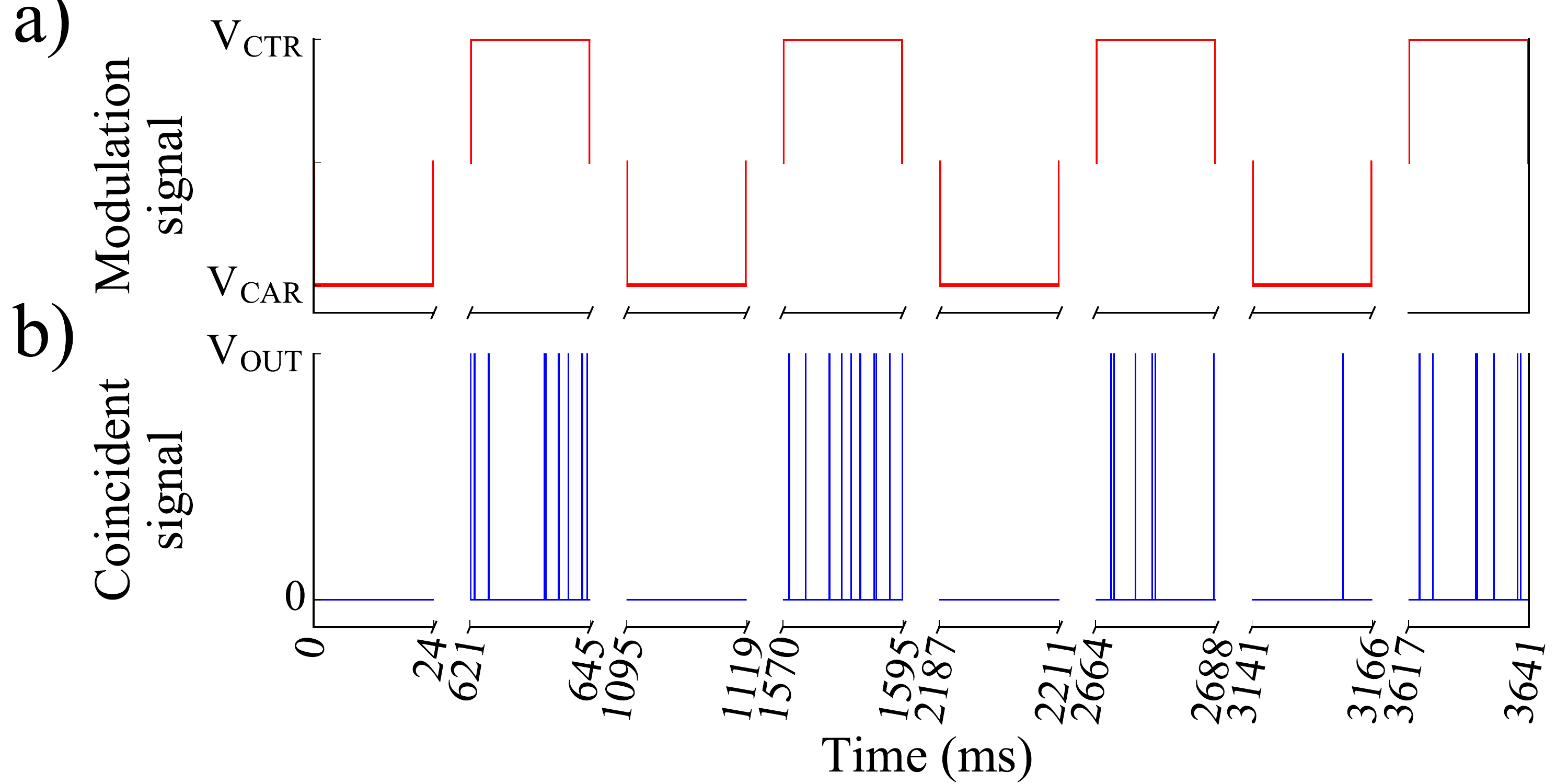}\\
\includegraphics[width=\columnwidth]{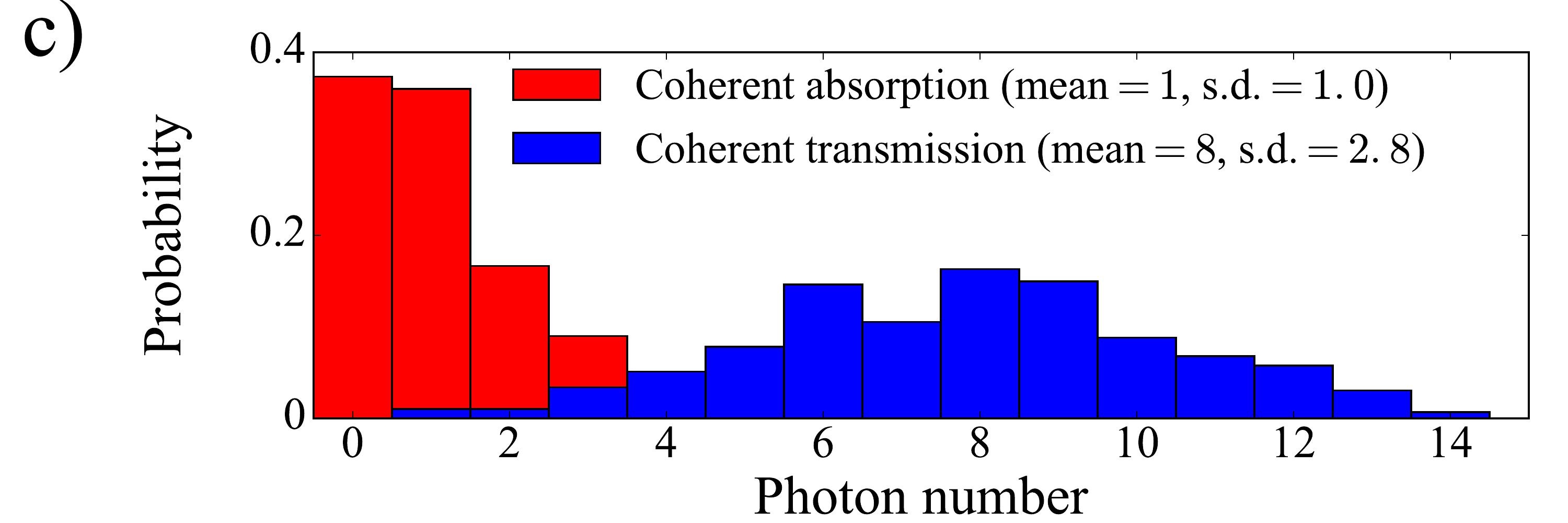}
\caption{\label{fig:4}\small Dissipative single-photon switching. 
The raw data: a) The modulation signal applied to the fibre stretcher driving the system between coherent absorption (CAR) and transmission regimes (CTR). b) The coincident photon detection signal between  SPD-c \& SPD-h and SPD-d \& SPD-h. ((a) and (b) share the x-axis, of which broken parts correspond to phase stabilization periods.) c) The coincidence counts distribution of coherent absorption (red) and tranmsission (blue) cycles.
}
\end{figure}

\setlength{\parindent}{3ex}Similar to the previous experiment, Fig.~\ref{fig:1}\textcolor{blue}{(c)}, we use two independent single-photon sources: 1) a strongly attenuated CW laser at a wavelength of 810 nm, and 2) a heralded single-photon source based on degenerate spontaneous parametric down-conversion (SPDC) in a non-linear crystal (BBO) pumped by CW laser at a wavelength of 405 nm. We use attenuated laser source to measure the interference fringes, and we use the heralded source to demonstrate single-photon switching beyond the dark count noise. In the latter case, the detection of idler photon of SPDC source by the single-photon detector SPD-h heralds the presence of the signal photon, which is sent to optical fibre network through a 50:50 BS. The second input port of the BS is used to launch photons from the attenuated laser. The coherent optical fibre network is comprised of a MZI where a delay line (to equal the interferometer arms within $100~\mu$m of the coherence length of the heralded photons) and a  fibre stretcher (used as a phase modulator) are inserted in the bottom arm, and a variable attenuator is placed in the upper arm to equal the losses in the optical paths (not shown for simplicity). To separate photons propagating in different directions, circulators are used in both arms of the interferometer. After splitting on the first input 50:50 BS, the photon is recombined in the middle of the network, where fully-fiberized coherent perfect absorber is placed. To fabricate the coherent absorber with the desired parameters~(Eq.~\ref{eq:CPA-parameters}), we exploit split-ring resonator structure manufactured on a 50 nm thick gold film deposited on the end-facet of the optical fibre (for details see~\cite{Anton:19,doi:10.1063/1.5040829}).

In Fig.~\ref{fig:1}\textcolor{blue}{(c)}, the length of each arm of the interferometer is around 20 m. The phase noise is similar to the one shown in Fig.~\ref{fig:2}\textcolor{blue}{(a)}, thus we keep $\Delta t=24$~ms. Since the outputs of the absorber are expected to behave in phase (see~Eq.~\ref{eq:CPA-p}), we use the total single-photon counts of both SPD-c ($N_\textrm{c}$) and SPD-d ($N_\textrm{d}$) measured during $\Delta t$,

\begin{equation}
\label{eq:CPA-N}
N_\textrm{c} + N_\textrm{d} = N \cdot (1 - \cos \phi)/2
\end{equation}

\setlength{\parindent}{0ex}as a reference signal for the stabilization. Here $N$ is the number of photons impinging on the absorber during $\Delta t$. 

\setlength{\parindent}{3ex}The interference fringes are measured following the stabilization of the system. Our experimental results show that $N_\textrm{c}$ and $N_\textrm{d}$ oscillate almost in phase (with a shift of $\pi/3$), which is in a good agreement with the expected results according to~Eq.~(\ref{eq:CPA-p}-\ref{eq:CPA-N}). The $\pi/3$ phase shift is due to the imperfections in the device fabrication and in-phase behaviour can be obtained by fabricating the metasurface symmetrically or by using a matching gel inside the metadevice package~\cite{Anton:19}.

During a $2\pi$-phase scan, the system passes the regimes of coherent absorption (minimum of the $N_\textrm{c}+N_\textrm{d}$ counts, $N_\textrm{min}$) and coherent transmission (maximum of the $N_\textrm{c}+N_\textrm{d}$, $N_\textrm{max}$) with visibility,

\begin{equation}
(N_\textrm{max} - N_\textrm{min})/(N_\textrm{max} + N_\textrm{min})
\end{equation}

\setlength{\parindent}{0ex}of 73\%. This visibility is lower than the visibility of individual curves (89\% for  $N_\textrm{c}$ and 86\% for $N_\textrm{d}$) due to the aforementioned nonideality of the sample and the unity system visibility is achievable if the device nonideality is mitigated. 

\setlength{\parindent}{3ex}This result is close to the one demonstrated previously~\cite{Anton:19}. Nonetheless, thanks to the new stabilization scheme, now we are able to control the absorption probability on-demand, which is crucial for practical applications of CPA phenomenon in quantum light processing. 
In this regard, we implement a dissipative single-photon switch as a proof of principle application.
Here, we use the heralded single-photon source for data acquisition to overcome the dark count noise. 
After each stabilization cycle, the system is driven to either the coherent absorption or transmission regime (Fig.~\ref{fig:4}\textcolor{blue}{(a)}), and the coincidence counts of SPD-c \& SPD-h and SPD-d \& SPD-h are recorded during $\Delta t$, Fig.~\ref{fig:4}\textcolor{blue}{(b)}. The flux of photons passes through the absorber at coherent transmission regimes, while it is almost totally absorbed during coherent absorption regimes. The photon distributions (Fig.~\ref{fig:4}\textcolor{blue}{(c)}) correspond to Poisson statistics of randomly generated heralded photons, where the data is acquired from 300 transmission and absorption cycles. On average,  8 photons with a standard deviation (s.d.) 2.8 are detected during the transmission cycle and 1 photon with a s.d. 1.0 is detected during the absorption cycle, with a switching visibility of 78\%. 

In summary, we showed a simple yet powerful technique of phase stabilization of coherent optical fibre networks based on single-photon counting. The method is able to overcome phase noise inherent in optical fibre systems with no need for auxiliary laser source and additional optical components required in conventional approaches. Achieved phase stability of 0.07 rad allowed us to control single-photon absorption on-demand in a fully fiberized MZI. Moreover, the method applied to a quantum network operating in the regime of the CPA made possible to realize dissipative switching at the single-photon level with visibility close to 80\%. Significant hardware simplification brought about by the proposed scheme is promising. Moreover, the development of relevant quantum technologies such as single-photon detectors with a decreased dead time, bright sources of quantum light and high-performance integrated optics would allow the real-world applications of this technique in coherent quantum communication and computation systems. 

\medskip
{\Large\bfseries\sffamily Funding.} 
Singapore A*STAR QTE program (SERC A1685b0005); Singapore Ministry of Education (MOE2016-T3-1-006 (S)); UK's Engineering and Physical Sciences Research Council (grant EP/M009122/1).

\medskip
{\Large\bfseries\sffamily Acknowledgment.} 
The authors thank Victor Leong and Rainer Dumke for technical discussions.

\small 
\bibliography{master}

\begin{thebibliography}{10}
\newcommand{\enquote}[1]{``#1''}

\bibitem{Xavier:11}
G.~B. Xavier and J.~P. von~der Weid, \enquote{Stable single-photon interference
  in a 1 km fiber-optic mach--zehnder interferometer with continuous phase
  adjustment,} {\protect\JournalTitle{Opt. Lett.}} \textbf{36}, 1764--1766
  (2011).

\bibitem{Cho:09}
S.-B. Cho and T.-G. Noh, \enquote{Stabilization of a long-armed fiber-optic
  single-photon interferometer,} {\protect\JournalTitle{Opt. Express}}
  \textbf{17}, 19027--19032 (2009).

\bibitem{Lukens:17}
J.~M. Lukens and P.~Lougovski, \enquote{Frequency-encoded photonic qubits for
  scalable quantum information processing,} {\protect\JournalTitle{Optica}}
  \textbf{4}, 8--16 (2017).

\bibitem{Lu:18}
H.-H. Lu, J.~M. Lukens, N.~A. Peters, B.~P. Williams, A.~M. Weiner, and
  P.~Lougovski, \enquote{Quantum interference and correlation control of
  frequency-bin qubits,} {\protect\JournalTitle{Optica}} \textbf{5}, 1455--1460
  (2018).

\bibitem{Li:18}
D.-D. Li, S.~Gao, G.-C. Li, L.~Xue, L.-W. Wang, C.-B. Lu, Y.~Xiang, Z.-Y. Zhao,
  L.-C. Yan, Z.-Y. Chen, G.~Yu, and J.-H. Liu, \enquote{Field implementation of
  long-distance quantum key distribution over aerial fiber with fast
  polarization feedback,} {\protect\JournalTitle{Opt. Express}} \textbf{26},
  22793--22800 (2018).

\bibitem{Liu:18}
H.~Liu, J.~Wang, H.~Ma, and S.~Sun, \enquote{Polarization-multiplexing-based
  measurement-device-independent quantum key distribution without phase
  reference calibration,} {\protect\JournalTitle{Optica}} \textbf{5}, 902--909
  (2018).

\bibitem{Sun:17}
Q.-C. Sun, Y.-F. Jiang, Y.-L. Mao, L.-X. You, W.~Zhang, W.-J. Zhang, X.~Jiang,
  T.-Y. Chen, H.~Li, Y.-D. Huang, X.-F. Chen, Z.~Wang, J.~Fan, Q.~Zhang, and
  J.-W. Pan, \enquote{Entanglement swapping over 100 km optical fiber with
  independent entangled photon-pair sources,} {\protect\JournalTitle{Optica}}
  \textbf{4}, 1214--1218 (2017).

\bibitem{Ikuta2018}
T.~Ikuta and H.~Takesue, \enquote{Four-dimensional entanglement distribution
  over 100{\hspace{0.167em}}km,} {\protect\JournalTitle{Scientific Reports}}
  \textbf{8} (2018).

\bibitem{Roger2015}
T.~Roger, S.~Vezzoli, E.~Bolduc, J.~Valente, J.~J.~F. Heitz, J.~Jeffers,
  C.~Soci, J.~Leach, C.~Couteau, N.~I. Zheludev, and D.~Faccio,
  \enquote{Coherent perfect absorption in deeply subwavelength films in the
  single-photon regime,} {\protect\JournalTitle{Nature Communications}}
  \textbf{6} (2015).

\bibitem{PhysRevLett.117.023601}
T.~Roger, S.~Restuccia, A.~Lyons, D.~Giovannini, J.~Romero, J.~Jeffers,
  M.~Padgett, and D.~Faccio, \enquote{Coherent absorption of n00n states,}
  {\protect\JournalTitle{Phys. Rev. Lett.}} \textbf{117}, 023601 (2016).

\bibitem{Anton:19}
A. N. Vetlugin, R. Guo, A. Xomalis, S. Yanikgonul, G. Adamo, C. Soci and N. I.
  Zheludev, "Coherent perfect absorption of single photons in a fibre network,"
  Appl. Phys. Lett, doc. ID APL19-AR-06056R1 (posted 23 October 2019, in
  press).

\bibitem{Cho:16}
S.-B. Cho and H.~Kim, \enquote{Active stabilization of a fiber-optic two-photon
  interferometer using continuous optical length control,}
  {\protect\JournalTitle{Opt. Express}} \textbf{24}, 10980--10986 (2016).

\bibitem{Musha:82}
T.~Musha, J.~ichi Kamimura, and M.~Nakazawa, \enquote{Optical phase
  fluctuations thermally induced in a single-mode optical fiber,}
  {\protect\JournalTitle{Appl. Opt.}} \textbf{21}, 694--698 (1982).

\bibitem{doi:10.1119/1.1286663}
E.~D. Black, \enquote{An introduction to pound–drever–hall laser frequency
  stabilization,} {\protect\JournalTitle{American Journal of Physics}}
  \textbf{69}, 79--87 (2001).

\bibitem{doi:10.1063/1.5040829}
A.~Xomalis, I.~Demirtzioglou, Y.~Jung, E.~Plum, C.~Lacava, P.~Petropoulos,
  D.~J. Richardson, and N.~I. Zheludev, \enquote{Picosecond all-optical
  switching and dark pulse generation in a fibre-optic network using a
  plasmonic metamaterial absorber,} {\protect\JournalTitle{Applied Physics
  Letters}} \textbf{113}, 051103 (2018).

\end{thebibliography}
\bibliographyfullrefs{master}
\end{document}